\documentclass[twocolumn,fullpaper]{jpsj2}

\def\runauthor{M. Matsumoto and M. Sigrist}

\title{Ehrenfest Relations and Magnetoelastic Effects in Field-Induced Ordered Phases}

\author{Masashige Matsumoto and Manfred Sigrist$^1$}
\inst{Department of Physics, Shizuoka University, Shizuoka 422-8529 \\
$^1$Theoretische Physik, ETH-H\"{o}nggerberg, CH-8093, Z\"{u}rich, Switzerland}

\recdate{March 11, 2005}
\abst{
Magnetoelastic properties in field-induced magnetic ordered phases
are studied theoretically based on a Ginzburg-Landau theory.
A critical field for the field-induced ordered phase is obtained
as a function of temperature and pressure,
which determine the phase diagram.
It is found that magnetic field dependence of elastic constant
decreases discontinuously at the critical field, $H_c$,
and that it decreases linearly with field in the ordered phase ($H>H_c$).
We found an Ehrenfest relation between the field dependence of the elastic constant
and the pressure dependence of critical field.
Our theory provides the theoretical form for magnetoelastic properties
in field- and pressure-induced ordered phases.
}

\kword{
magnetoelasticity,
quantum phase transition,
elastic constant,
thermal expansion,
spin dimer
}

\begin{document}
\sloppy
\maketitle

%%%%%%%%%%%%%%%%%%%%%%%%%%%%%%%%%%%%%%%%%%%%%%%%%%%%%%%%%%%%%%%%%%%%%%%%%%%%%%%%%%%%%%%%%%%%%%%%%%%
\newcommand{\br}{{\mbox{\boldmath$r$}}}
\newcommand{\bk}{{\mbox{\boldmath$k$}}}
\newcommand{\sk}{{\mbox{\footnotesize $k$}}}
\newcommand{\bsk}{{\mbox{\footnotesize \boldmath$k$}}}
\newcommand{\bsq}{{\mbox{\footnotesize \boldmath$q$}}}
\newcommand{\bsQ}{{\mbox{\footnotesize \boldmath$Q$}}}
\newcommand{\bsr}{{\mbox{\footnotesize \boldmath$r$}}}
\newcommand{\bS}{{\mbox{\boldmath$S$}}}
\newcommand{\bQ}{{\mbox{\boldmath$Q$}}}
\newcommand{\bd}{{\mbox{\boldmath$d$}}}
\newcommand{\bsd}{{\mbox{\footnotesize{\boldmath$d$}}}}
\newcommand{\bsigma}{{\mbox{\boldmath$\sigma$}}}
\newcommand{\Tl}{{TlCuCl$_3$}}
\newcommand{\K}{{KCuCl$_3$}}
\newcommand{\N}{{NH$_4$CuCl$_3$}}
\newcommand{\Sr}{{SrCu$_2$(BO$_3$)$_2$}}
\newcommand{\ha}{{\hat{a}}}
\newcommand{\hb}{{\hat{b}}}
\newcommand{\hc}{{\hat{c}}}
\newcommand{\hx}{{\hat{x}}}
\newcommand{\hy}{{\hat{y}}}
\newcommand{\hz}{{\hat{z}}}
\newcommand{\muB}{{\mu_{\rm B}}}
\newcommand{\del}{\partial}
%%%%%%%%%%%%%%%%%%%%%%%%%%%%%%%%%%%%%%%%%%%%%%%%%%%%%%%%%%%%%%%%%%%%%%%%%%%%%%%%%%%%%%%%%%%%%%%%%%%

%%%%%%%%%%%%%%%%%%%%%%%%%%%%%%%%%%%%%%%%%%%%%%%%%%%%%%%%%%%%%%%%%%%%%%%%%%%%%%%%%%%%%%%%%%%%%%%%%%%
\section{Introduction}
%%%%%%%%%%%%%%%%%%%%%%%%%%%%%%%%%%%%%%%%%%%%%%%%%%%%%%%%%%%%%%%%%%%%%%%%%%%%%%%%%%%%%%%%%%%%%%%%%%%

Lattices of interacting quantum spin $S=1/2$ dimers
provide a fascinating class of systems to study various aspects of quantum phase transitions.
\cite{Rice}
For weak interdimer interaction,
the dimer spins lock into spin singlets and have only short range correlations.
As these interactions grow, a transition to a magnetically long range ordered state can appear, 
depending on the coupling topology.
The dimer structure allows a useful representation of the spin degrees of freedom 
in terms of dimer basis: one spin singlet and three spin triplet states.
In the disordered phase, the unique ground state
is adiabatically connected with the simple product of dimer singlets.
The elementary excitations are gapped magnons and correspond to dimer spin triplets.
These magnons are hard-core bosons which acquire a finite dispersion,
because they can propagate through the lattices by the interdimer coupling.
The quantum phase transition to the magnetically ordered state
coincides with the closing of the excitation gap.
Hence, the ordering transition can be described
as a Bose-Einstein condensation (BEC) of magnons.
\cite{Oosawa-1999,Nikuni}

The quantum phase transition can be induced
through the variation of some external parameters.
One is external magnetic field
which leads to a splitting of the spin triplet multiplet by Zeeman coupling.
\cite{Tanaka-1998,Cavadini-2002}
The lowest triplet state removes the gap at a critical magnetic field
and gives rise to an ordered phase.
\cite{Ruegg-2003,Matsumoto-2002}
Another parameter is pressure which affects the interdimer coupling,
in general, strengthening them such that the magnon dispersion increases
and the gap eventually closes at a critical pressure.
\cite{Oosawa-2003,Ruegg-2004-pressure,Matsumoto-2004}

There are various ways to investigate
the phases and phase transitions in these spin dimer systems
like neutron scattering, electron spin resonance, nuclear magnetic resonance (NMR)
and magnetization measurements.
In addition to these, also thermodynamic properties are important
to understand the phase transition of these systems.
Ultrasound experiments and the thermal expansion
are the most easily accessible properties
and allow us to probe the phase transitions.
Schmidt {\it et al}. have measured the field dependence of the elastic constant of \Sr~and \N.
They found a discontinuity in the field dependence of the elastic constant at the transition
with a linear field dependence in the ordered phase.
\cite{Schmidt,Luthi,Wolf}
They discuss that the field dependence of the elastic constants in the ordered phases
resembles very much that of the uniform differential susceptibility, $d M/d H$.
The thermal expansion in \Tl, measured by Johannsen {\it et al}.,
indicates a strong uniaxial pressure dependence of the magnon gap.
\cite{Johannsen}
However, there are few theories for the magnetoelastic effect
in the field- and pressure-induced ordered phases.

In this paper, we investigate magnetoelastic properties of the field- and pressure-induced orders,
focusing on copper-chrolides \Tl, \K~and \N.
We will introduce a phenomenological formulation for the magnetoelastic effects
which is the basis of our discussion of thermodynamic properties
and the Ehrenfest relations at the phase transitions.
The spin-lattice coupling relevant for specific materials
will be also discussed from microscopic point of view.

%%%%%%%%%%%%%%%%%%%%%%%%%%%%%%%%%%%%%%%%%%%%%%%%%%%%%%%%%%%%%%%%%%%%%%%%%%%%%%%%%%%%%%%%%%%%%%%%%%%
\section{Ginzburg-Landau theory}
%%%%%%%%%%%%%%%%%%%%%%%%%%%%%%%%%%%%%%%%%%%%%%%%%%%%%%%%%%%%%%%%%%%%%%%%%%%%%%%%%%%%%%%%%%%%%%%%%%%

This section is devoted to the formulation of a Ginzburg-Landau (GL) theory
to describe magnetoelastic effects in the vicinity of the quantum phase transitions.
Such theories have been developed in different context previously.
\cite{Walker,Plumer}
We extend the Ginzburg-Landau expansion of the free energy
to the case of field- and pressure-induced (quantum) phase transitions.
The order parameter describing the spontaneously broken symmetry
is the staggered magnetic moment $m$ which lies perpendicular to the applied magnetic field.
\cite{Tanaka-2001}
We require that this theory gives a description of the basic phase diagram
in temperature, magnetic field and pressure as we know it to some extent for \Tl.
The free energy density for the field- and pressure-induced ordering is given by
\begin{align}
F = A m^2 + B m^4 + \frac{1}{2}c_0\epsilon^2 + \epsilon P
  - K \epsilon \left( \frac{T}{T_0} \right)^a,
\label{eqn:freeenergy}
\end{align}
where the first and the second terms constitute the magnetic free energy.
The latter three terms describe the lattice with $\epsilon$ as the strain
which may represent the volume or a uniaxial deformation
from the equilibrium position in the absence of an external pressure $P$.
$c_0$ is the corresponding elasticity constant,
and the last term describes the thermal expansion of the lattice.
$T_0$ is a characteristic temperature which can be determined by experiments.
The coefficient $B$ is a positive constant and $A =0$ determines the phase boundary:
\begin{align}
A = a_0 \left[ 1 + \left( \frac{T}{T_0} \right)^\phi - \left( \frac{H}{H_0} \right)^2 \right]
  + \gamma_1 \epsilon + \gamma_2 \epsilon^2,
\label{eqn:A}
\end{align}
with $a_0 > 0$ constant.
$H_0$ is a characteristic magnetic field which can be determined by experiments.
As we will see later, $H_0$ is the critical field for zero temperature and pressure.
The field dependence in $A$ is quadratic to lowest order
due to time reversal symmetry of the disordered phase.
The coupling to the lattice is given by an expansion in $\epsilon$.
Note that in order to achieve pressure induced order, $\gamma_1 > 0$ is required,
since under pressure $\epsilon$ is negative ($\epsilon < 0$).
The exponents $a$ and $\phi$ of $T$ in eqs. (\ref{eqn:freeenergy}) and (\ref{eqn:A})
have to be chosen to satisfies Nernst's theorem.
Since the entropy $S = - dF/dT$ and the thermal expansion $\alpha_1=\del \epsilon/\del T$
should vanish at $ T=0$, we find the conditions: $a > 1$ and $\phi > 1$.

%%%%%%%%%%%%%%%%%%%%%%%%%%%%%%%%%%%%%%%%%%%%%%%%%%%%%%%%%%%%%%%%%%%%%%%%%%%%%%%%%%%%%%%%%%%%%%%%%%%
\subsection{Basic magnetic properties}
%%%%%%%%%%%%%%%%%%%%%%%%%%%%%%%%%%%%%%%%%%%%%%%%%%%%%%%%%%%%%%%%%%%%%%%%%%%%%%%%%%%%%%%%%%%%%%%%%%%

The free energy has to be minimized with respect to $m$ and $\epsilon$
for given temperature $T$, external field $H$ and pressure $P$.
First, we assume for simplicity that the pressure determines $\epsilon$.
We determine the staggered moment $m$ and the uniform magnetization $M$
by the following equations:
\begin{align}
\frac{\del F}{\del m^2} = 0,~~~
M = -\frac{\del F}{\del H} = \frac{2 a_0 m^2}{H_0} \frac{H}{H_0},
\end{align}
which lead to 
\begin{align}
m^2 \simeq
\left\{
\begin{array}{@{\,}lr}  \displaystyle
  \frac{a_0}{2B} \frac{2H_c\delta H+\delta H^2}{H_0^2} & P \leq P_c,~H \geq H_c
  \\ & \\  \displaystyle
  \frac{1}{2B} \left[ |a_{0T}| + a_0 \left( \frac{H}{H_0} \right)^2 \right] & P\geq P_c,~H \geq 0
\end{array}
\right. ,
\label{eqn:exponent} 
\end{align}
and
\begin{align}
M \simeq
\left\{
\begin{array}{@{\,}lr} \displaystyle
  \frac{a_0^2}{B H_0}\frac{2H_c^2\delta H + 3H_c\delta H^2 + \delta H^3}{H_0^3}
  & P\leq P_c,~H \geq H_c
  \\ & \\   \displaystyle
  \frac{a_0}{B H_0} \left[ |a_{0T}| \frac{H}{H_0}
                          + a_0 \left( \frac{H}{H_0} \right)^3 \right] & P\geq P_c,~H \geq 0
\end{array}
\right. , 
\label{uniform-mag}
\end{align}
where
\begin{align}
a_{0T} &= a_0 \left[ 1 + \left( \frac{T}{T_0} \right)^\phi \right]
        + \gamma_1 \epsilon + \gamma_2 \epsilon^2, \cr
\delta H &= H - H_c, \cr
H_c &= H_c(T,P) = H_0 \sqrt{\frac{a_{0T}}{a_0}}.
\label{eqn:Hc-define}
\end{align}
Here, $P_c$ and $H_c$ represent critical field and pressure, respectively.
Depending on the applied pressure (or strain),
$a_{0T}$ is either positive ($ P < P_c $) or negative ($P>P_c$).
The first case ($ P < P_c $) leads to field-induce order,
when $H$ exceeds the critical field $H_c$.
Here we find $M \propto H - H_c$ and $m \propto (H-H_c)^{1/2}$.
For the second case ($ P > P_c $), order appears for zero magnetic field,
giving the following field dependences:
$M \propto H$ and $m - m_0 \propto H^2 $,
where $m_0 = (|a_{0T}|/2B)^{1/2}$ is the staggered moment at $H=0$.
At the boundary between the two cases, $P=P_c$,
the critical field is reduced to zero $H_c(T,P_c)=0$,
since $a_{0T}=0$ in eq. (\ref{eqn:Hc-define}).
At the critical pressure, we find $M \propto H^3$ and $m \propto H$.
These field dependences in eqs. (\ref{eqn:exponent}) and (\ref{uniform-mag})
(exponents listed in Table I) do not depend on the GL parameters
($H_0$, $T_0$, $P_0$, $\phi$ and $a$),
and are consistent with the results
obtained from the bond-operator theory for the interacting spin dimer system.
\cite{Matsumoto-2004}
Thus, the GL theory with free energy (\ref{eqn:freeenergy})
is equivalent to the bond-operator theory in the vicinity of the phase transitions
and otherwise at least useful for a qualitative analysis.
Very recently, Goto {\it et al}. performed magnetization measurements under pressure
and confirmed that the uniform magnetization actually behaves as $M \sim H^3$ at $P=P_c$.
\cite{Goto}
We note that the recent Monte Carlo simulations by Nohadani {\it et al}.
also support these exponents.
\cite{Nohadani-preprint}

%%%%%%%%%%%%%%%%%%%%%%%%%%%%%%%%%%%%%%%%%%%%%%%%%%%%%%%%%%%%%%%%%%%%%%%%%%%%%%%%%%%%%%%%%%%%%%%%%%%
\begin{table}[t]
\caption{
Exponents for the staggered ($m$) and uniform ($M$) magnetizations obtained by the GL theory.
$m_0$ is the induced staggered moment at $H=0$ above the critical pressure.
}
\begin{displaymath}
\begin{array}{l|l|l} \hline
P < P_c     & P = P_c & P > P_c \\ \hline
m \propto (H-H_c)^{\frac{1}{2}} & m \propto H & m-m_0 \propto H^2 \\
M \propto H-H_c & M \propto H^3 & M \propto H \\ \hline
\end{array}
\end{displaymath}
\end{table}
%%%%%%%%%%%%%%%%%%%%%%%%%%%%%%%%%%%%%%%%%%%%%%%%%%%%%%%%%%%%%%%%%%%%%%%%%%%%%%%%%%%%%%%%%%%%%%%%%%%

%%%%%%%%%%%%%%%%%%%%%%%%%%%%%%%%%%%%%%%%%%%%%%%%%%%%%%%%%%%%%%%%%%%%%%%%%%%%%%%%%%%%%%%%%%%%%%%%%%%
\subsection{Phase diagram}
%%%%%%%%%%%%%%%%%%%%%%%%%%%%%%%%%%%%%%%%%%%%%%%%%%%%%%%%%%%%%%%%%%%%%%%%%%%%%%%%%%%%%%%%%%%%%%%%%%%

\begin{figure}[t]
\begin{center}
\includegraphics[width=7cm]{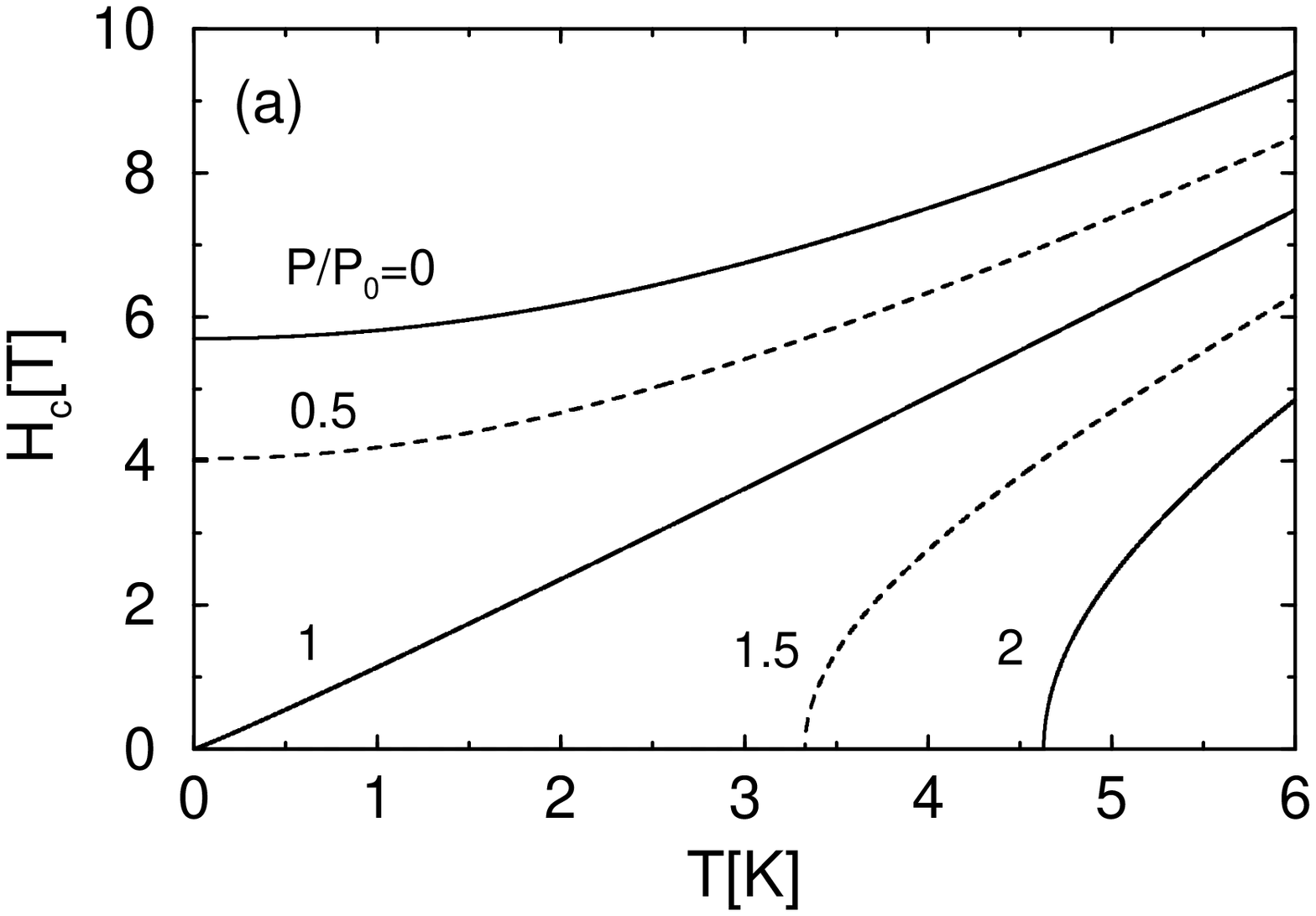}
\includegraphics[width=7cm]{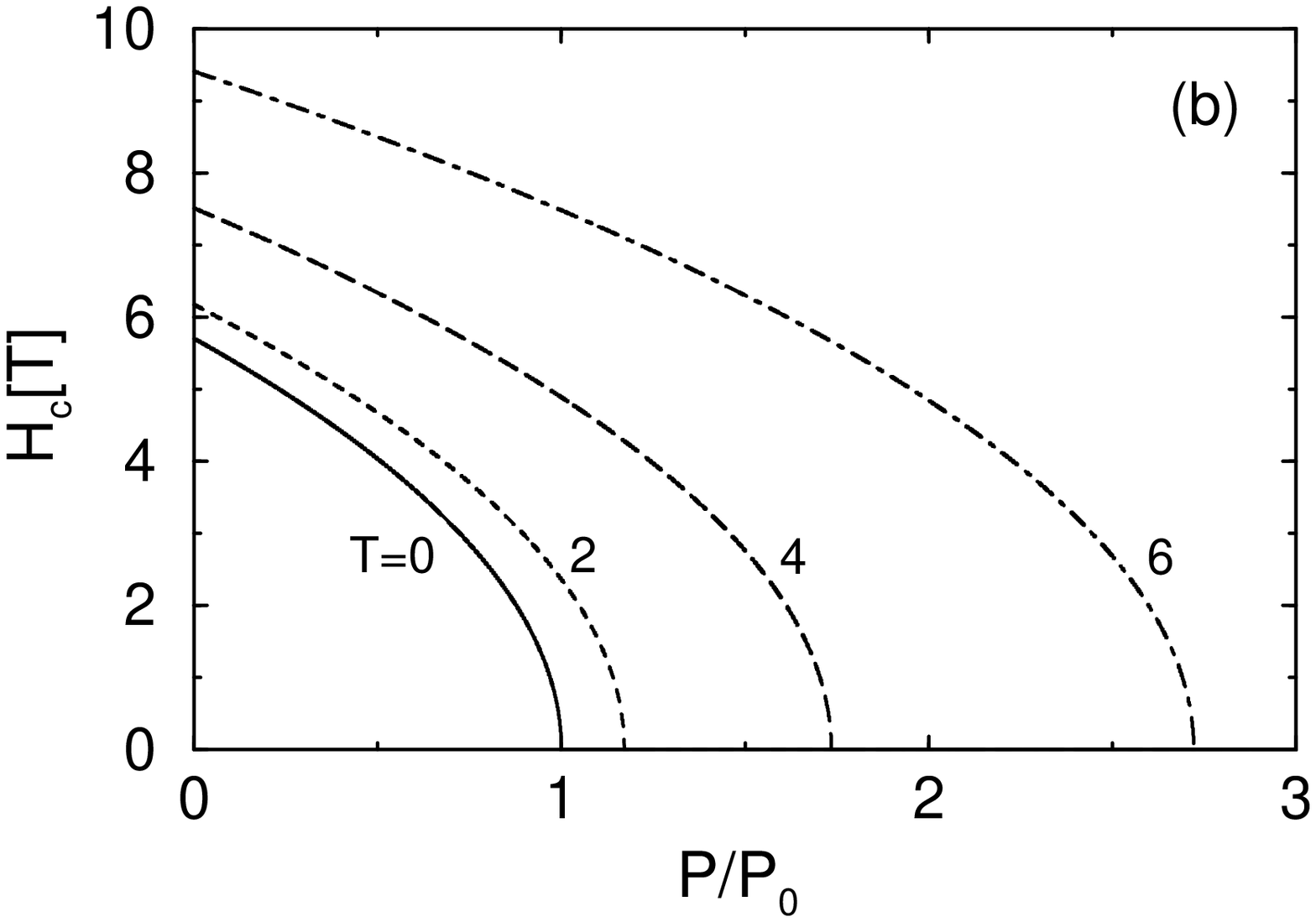}
\includegraphics[width=6.8cm]{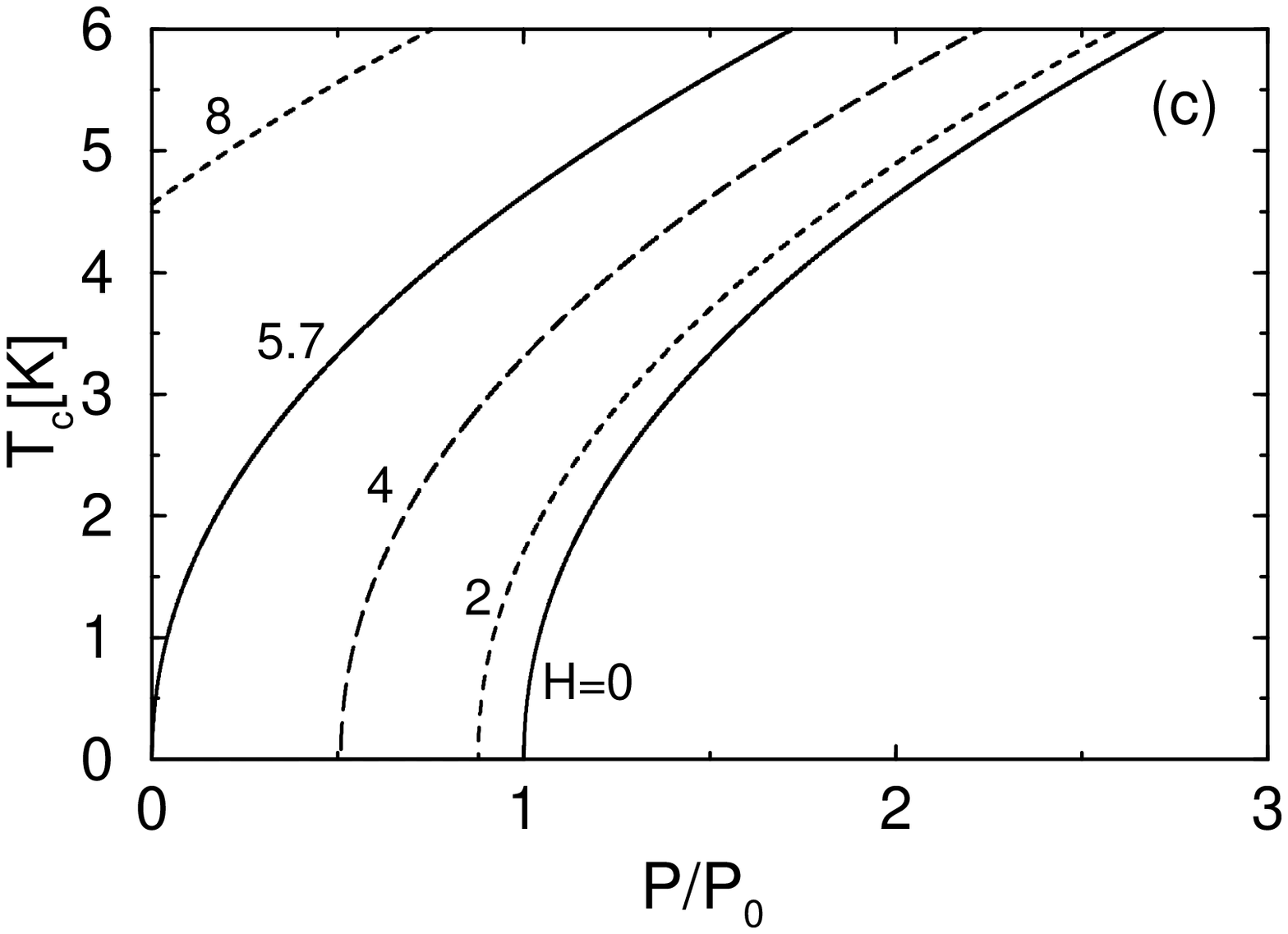}
\end{center}
\caption{
(a) $T-H$ phase diagram for fixed pressure.
Values of pressure are on the figure in unit of the critical pressure $P_0$.
(b) $P-H$ phase diagram for fixed temperature.
Values of temperature are on the figure in unit of Kelvin.
(c) $P-T$ phase diagram for fixed magnetic field.
Values of magnetic field are on the figure in unit of Tesla.
Parameters are chosen as $H_0=5.7$ T, $\phi=2.1$ and $T_0=4.6$ K.
}
\label{fig:phase}
\end{figure}

In this subsection, we demonstrate that the phase diagram is obtained
by giving appropriate values for the GL parameters $H_0$, $T_0$, $P_0$ and $\phi$.
As mentioned above, the phase boundary as a function of temperature, field and pressure
is determined by the vanishing second-order coefficient $A$.
We assume that the system is disordered for $T=0$, $H=0$ and $P=0$.
Considering magnetoelastic aspects of the phase diagram,
we neglect for the moment the term $\gamma_2 \epsilon^2$.
Hence, the critical pressure is easily obtained
by minimizing first the lattice part of the free energy with respect to $\epsilon$,
which leads to
\begin{align}
\epsilon = - \frac{P}{c_0}.
\end{align}
From this, the critical pressure for zero temperature and field is obtained as
\begin{align}
P_c(0,0) = P_c(T=0,H=0) = \frac{a_0 c_0}{\gamma_1} = P_0.
\end{align}
Here, $P_0$ can be determined by experiments.
The critical field for zero temperature and pressure is given by
\begin{align}
H_c(0,0) = H_c(T=0,P=0) = H_0.
\end{align}
Therefore, $H_0$ introduced in eq. (\ref{eqn:A})
is the critical field for $T=0$ and $P=0$.
In the finite ($T,P,H$)-diagram, the boundary is determined by
\begin{align}
\left(\frac{H}{H_0} \right)^2 + \frac{P}{P_0} - \left( \frac{T}{T_0} \right)^\phi = 1.
\label{phas-bound}
\end{align}
The critical field, temperature and pressure are expressed as
\begin{align}
H_c(T,P) &= H_0 \sqrt{ 1 + \left( \frac{T}{T_0} \right)^\phi - \frac{P}{P_0} }, \cr
T_N(H,P) &= T_0 \left[ \left( \frac{H}{H_0} \right)^2
                     + \frac{P}{P_0} -1 \right]^\frac{1}{\phi}, \cr
P_c(T,H) &= P_0 \left[ 1 + \left( \frac{T}{T_0} \right)^\phi
                         - \left( \frac{H}{H_0} \right)^2 \right].
\label{eqn:Hc}
\end{align}
The parameters $H_0$, $P_0$, $T_0$ and $\phi$
can be chosen as to reproduce the experimentally given phase diagram.
In Fig.\ref{fig:phase}, the phase boundary lines are shown
as obtained from the GL theory with the basic parameters found for \Tl.
The three figures provide cuts for two variables fixing the third one.

%%%%%%%%%%%%%%%%%%%%%%%%%%%%%%%%%%%%%%%%%%%%%%%%%%%%%%%%%%%%%%%%%%%%%%%%%%%%%%%%%%%%%%%%%%%%%%%%%%%
%%%%%%%%%%%%%%%%%%%%%%%%%%%%%%%%%%%%%%%%%%%%%%%%%%%%%%%%%%%%%%%%%%%%%%%%%%%%%%%%%%%%%%%%%%%%%%%%%%%

At atmospheric pressure ($P=0$), the phase boundary has the form
\begin{align}
H_c(T,0) - H_0 &= H_0 \left[ \sqrt{1 + \left( \frac{T}{T_0} \right)^\phi} -1 \right] \cr
&\simeq \frac{1}{2} H_0 \left( \frac{T}{T_0} \right)^\phi.
\end{align}
The exponent $\phi$ can therefore be determined from the experimental phase boundary.
For \Tl, Oosawa {\it et al.} examined the phase diagram on $T-H$ plane
at the atmospheric pressure by specific heat measurements,
which give the values $H_0 \simeq 5.7$ T, $\phi \simeq 2.1$, and $T_0 \simeq 4.6$ K.
\cite{Oosawa-2001}
Note that the exponent satisfies safely the condition $\phi > 1$ given above.

The critical pressure at $T=0$ and $H=0$ is determined by Goto {\it et al}.
by measuring the magnetization curve: $P_0 \simeq 0.4$ kbar,
\cite{Goto}
while inelastic neutron scattering experiment under pressure by R\"{u}egg {\it et al}.
reported that $P_0 \simeq 1.0$ kbar.
\cite{Ruegg-2004-pressure}
The latter value was obtained by examining the pressure where the excitation gap closes.
There is discrepancy in the critical pressure between the two experiments,
and therefore, we leave $P_0$ as a parameter
and renormalize the pressure $P$ in Fig. \ref{fig:phase}.

The temperature dependence of $H_c(T,P)$ for fixed $P$ is shown in Fig. \ref{fig:phase}(a),
where the ordered phase lies above the phase boundaries.
Figure \ref{fig:phase}(b) displays the $P$ dependence of $H_c(T,P)$ for fixed $T$,
where magnetic order appears above the lines.
Increasing the temperature enlarges the disordered region.
Figure \ref{fig:phase}(c) shows the phase boundaries
in the $P-T$ plane for fixed magnetic fields.
The ordered phase lies below the lines.

The result of $\phi=2.1$ is in apparent conflict with the theories
discussing the universal properties of BEC
which leads to $\phi_{\rm BEC} = 3/2$.
%\cite{Giamarchi,Nikuni}
\cite{Nikuni}
However, the temperature range,
where the universal exponents are valid,
may be rather small as was found from extensive numerical simulations
by Nohadani {\it et al}.
\cite{Nohadani}
Very recently, Kawashima confirmed this point
by studying an effective $XY$ model with numerical simulations.
\cite{Kawashima}
Misguich {\it et al}. also examined this point
by the mean-field theory of hard core boson of excited triplet magnons
with realistic dispersion relations.
\cite{Misguich}
They reported that the critical field behaves as $H_c(T,0)-H_0 \sim T^{\frac{3}{2}}$
only at quite low temperatures
and that the $\phi = 2.1$ can be seen in a wide range of temperatures.

%%%%%%%%%%%%%%%%%%%%%%%%%%%%%%%%%%%%%%%%%%%%%%%%%%%%%%%%%%%%%%%%%%%%%%%%%%%%%%%%%%%%%%%%%%%%%%%%%%%
\section{Ehrenfest relations}
%%%%%%%%%%%%%%%%%%%%%%%%%%%%%%%%%%%%%%%%%%%%%%%%%%%%%%%%%%%%%%%%%%%%%%%%%%%%%%%%%%%%%%%%%%%%%%%%%%%

In the previous section, we examined the basic magnetic property and phase diagram,
and showed that the results are consistent
with the recent experiments and numerical calculations.
In this section, we study properties of thermal expansion and elastic constant
within the GL theory and derive Ehrenfest relations satisfied at the phase transition.
We note that the Ehrenfest relations do not depend on details of the parameters
$H_0$, $T_0$, $P_0$, $\phi$ and $a$ introduced for the GL free energy.
The Ehrenfest relations help us study the magnetoelastic effects
which take place at the field- and pressure-induced magnetic phase transitions.

%%%%%%%%%%%%%%%%%%%%%%%%%%%%%%%%%%%%%%%%%%%%%%%%%%%%%%%%%%%%%%%%%%%%%%%%%%%%%%%%%%%%%%%%%%%%%%%%%%%
\subsection{Thermal expansion}
%%%%%%%%%%%%%%%%%%%%%%%%%%%%%%%%%%%%%%%%%%%%%%%%%%%%%%%%%%%%%%%%%%%%%%%%%%%%%%%%%%%%%%%%%%%%%%%%%%%

So far, we have discussed the influence of pressure or lattice deformation
on the phase transition and the magnetic order.
The feedback of the magnetic order on the lattice has been ignored so far.
This aspect of magnetoelasticity shall be now included,
since it can be measured with high sensitivity
and provides helpful probes to characterize the system further,
even giving access to certain microscopic features as we will show below.

We examine this by minimizing the free energy with respect to $\epsilon$
in addition to the staggered moment $m$.
We obtain the following behavior:
\begin{align}
\epsilon = \left\{ \begin{array}{ll} \displaystyle \frac{K}{c_0} \left( \frac{T}{T_0} \right)^a
 & {\rm disordered} \\ \\ \displaystyle
 \frac{K}{c_0} \left( \frac{T}{T_0} \right)^a
 - \frac{a_0 \gamma_1}{2 B c_0} \frac{H^2-H_c^2}{H_0^2} & {\rm ordered} \\
\end{array} \right.
\label{eqn:e}
\end{align}
Here, we neglect the quadratic coupling term with $\gamma_2$
which does not affect the result qualitatively.
The lattice expands with temperature as described by the first term.
The strain shows an abrupt change in the temperature and field dependence
at the onset of magnetic order.
For a fixed temperature, the magnetoelastic effect appears
as a linear field dependence in $\epsilon$ close to the phase transition,
since $H^2-H_c^2 \simeq 2H_c(H-H_c)$.
With increasing pressure, $H_c$ decreases as in eq. (\ref{eqn:Hc})
and becomes zero at the critical pressure for a given temperature.
In this case, the strain shows a quadratic field dependence,
since $H_c=0$ in eq. (\ref{eqn:e}).

The first derivative of $\epsilon$ in both $H$ and $T$ is discontinuous.
We derive the thermal expansion from the strain $\epsilon$ by
\begin{align}
\alpha_1 = \frac{\del \epsilon}{\del T},
\end{align}
which leads to
\begin{align}
\alpha_1 = \left\{ \begin{array}{ll}
   \displaystyle \frac{aK}{c_0 T_0} \left( \frac{T}{T_0} \right)^{a-1}
   & \mbox{disordered} \\ & \\ \displaystyle
\frac{aK}{c_0 T_0} \left( \frac{T}{T_0} \right)^{a-1}
   + \frac{a_0 \gamma_1}{2B c_0 H_0^2}
  \left( \frac{\partial H_c^2}{\partial T} \right)_P &\mbox{ordered}
\end{array} \right.
\end{align}
We find that $\alpha_1$ vanishes at zero temperature
for the given conditions on the exponents $a,~\phi > 1$.
At the N\'{e}el temperature $T_N$, we find a discontinuity in $\alpha_1$
which we can bring into relation with other thermodynamic quantities.
For this purpose, we express some of the phenomenological parameters
in the GL free energy by measurable quantities.
The coupling between strain and order parameter is connected
with the pressure dependence of the N\'{e}el temperature and the critical field.
Thus, a simple calculation results in
\begin{align}
\gamma_1 &= - \frac{a_0 c_0}{H_0^2} \left( \frac{\partial H_c^2}{\partial P} \right)_T
          = \frac{a_0 c_0}{T_0^\phi}\left( \frac{\partial T_N^\phi}{\partial P} \right)_H, \cr
\gamma_2 &= \frac{a_0 c_0^2}{2 H_0^2} \left( \frac{\del^2 H_c^2}{\del P^2} \right)_T
          = - \frac{a_0 c_0^2}{2 T_0^\phi}
              \left( \frac{\del^2 T_N^\phi}{\del P^2} \right)_H.
\label{eqn:gamma}
\end{align}

The entropy is given by
\begin{align}
S &= - \frac{dF}{dT} = - ( \frac{\partial F}{\partial T}
                + \frac{\partial F}{\partial m^2} \frac{d m^2}{d T}
                + \frac{\partial F}{\partial \epsilon} \frac{d \epsilon}{d T} )
= - \frac{\partial F}{\partial T} \cr
&= - m^2 \frac{\partial A}{\partial T}
   + \frac{a K \epsilon}{T_0} \left( \frac{T}{T_0} \right)^{a-1}.
\label{eqn:entropy}
\end{align}
Here, we used the following conditions to minimize the free energy:
$\partial F / \partial m^2 = 0$ and $\partial F / \partial \epsilon = 0$.
The specific heat is obtained by $C_p = T (\partial S / \partial T )_{H,P}$.
At the phase transition,
the first term in eq. (\ref{eqn:entropy}) leads to a discontinuity
in the specific heat with a value of
\begin{align}
\Delta C_p = \frac{a_0^2 \phi^2}{2 B T_0} \left( \frac{T_N}{T_0} \right)^{2 \phi -1} > 0.
\label{eqn:spec}
\end{align}
This value increases proportional to $ T_N^{2 \phi -1} $.
We may now express the ratio $a_0^2 / B$ in terms of $\Delta C_p$.

Now fix the magnetic field $H > H_0$
and consider the temperature dependence of the thermal expansion.
In the ordered phase, it is given by
\begin{align}
\alpha_1 = \frac{aK}{c_0 T_0} \left( \frac{T}{T_0} \right)^{a-1} + \frac{\Delta C_p}{T_N}
  \left( \frac{\partial T_N}{\partial P} \right)_H \left(\frac{T}{T_N} \right)^{\phi-1}
\label{eqn:alpha}
\end{align} 
which goes to zero as $T^{\phi-1}$ for $T \to 0$.
It is obtained by replacing the GL parameters by the calculated quantities
through eqs. (\ref{phas-bound}), (\ref{eqn:gamma}) and (\ref{eqn:spec}).
There is the following discontinuity of $\alpha_1$ at $T=T_N$:
\begin{align}
\Delta \alpha_1 = \frac{\Delta C_p}{T_N} \left( \frac{\partial T_N}{\partial P} \right)_H.
\label{eqn:ehrenfest-1}
\end{align}
This connection between discontinuities
is one of the Ehrenfest relations for the phase transition.

Analogously, we may consider the situation for fixed temperature
and increase the magnetic field to cross the phase boundary.
The corresponding expansion coefficient is defined as
\begin{align}
\alpha_2 = \frac{\partial \epsilon}{\partial H}
= \left\{ \begin{array}{ll} 0 & {\rm disordered} \\
  \displaystyle - \frac{a_0 \gamma_1}{B c_0 H_0}
                     \left( \frac{H_c}{H_0} + \frac{H-H_c}{H_0} \right) & {\rm ordered}
\end{array} \right. .
\end{align}
We note that $\alpha_2$ has a discontinuity at the transition ($H=H_c$)
and exhibits a linear field dependence in the ordered phase ($H>H_c$).
For varying field, the uniform magnetic moment $M$ and the susceptibility $\chi$
take roles formally equivalent to the entropy and specific heat, respectively.
\begin{align}
M = - \frac{\partial F}{\partial H}  \quad \mbox{and} \quad \chi = \frac{dM}{dH},
\end{align}
where $\chi$ has a discontinuity at the phase transition:
\begin{align}
\Delta \chi = \frac{2 a_0^2}{B H_0^2} \left( \frac{H_c}{H_0} \right)^2 > 0.
\label{eqn:chi}
\end{align}
Therefore, we find the following discontinuity at the transition for $\alpha_2$:
\begin{align}
\Delta \alpha_2 = - \frac{a_0 \gamma_1}{B c_0 H_0} \frac{H_c}{H_0}
                = \Delta \chi \left( \frac{\partial H_c}{\partial P} \right)_T,
\label{eqn:ehrenfest-2}
\end{align}
which is an Ehrenfest relation equivalent to eq. (\ref{eqn:ehrenfest-1}).
These Ehrenfest relations can give certain insights
into the microscopic understanding of the system.
In particular, the qualitative behavior of $T_N$ and $H_c$
as a function of uniform or uniaxial pressure can be guessed
by looking at the lattice structure and the involved exchange paths as we will discuss below.
The Ehrenfest relations can also be tested experimentally.
Recently, Johansen {\it et al}. found a behavior
consistent with the given Ehrenfest relations in \Tl~
for various uniaxial strains and corresponding expansion coefficients.
\cite{Johannsen}

Very recently, Sawai {\it et al}. reported that the uniform magnetization and strain
have quite similar field dependence in the field-induced ordered phase.
\cite{Sawai}
In the field-induced ordered phase,
we find the following relations in the vicinity of the critical field:
\begin{align}
M &= \Delta \chi ( H - H_c), \cr
\epsilon &= \frac{K}{c_0} \left( \frac{T}{T_0} \right)^a + \Delta \alpha_2 ( H - H_c) \cr
         &= \frac{K}{c_0} \left( \frac{T}{T_0} \right)^a
          + \Delta \chi \left( \frac{\partial H_c}{\partial P} \right)_T ( H - H_c).
\label{eqn:M-e}
\end{align}
Here, we used eqs. (\ref{eqn:chi}) and (\ref{eqn:ehrenfest-2}).
Both $M$ and $\epsilon$ have linear field dependence with related coefficients.
This relation (\ref{eqn:M-e}) can be tested by experiments.

%%%%%%%%%%%%%%%%%%%%%%%%%%%%%%%%%%%%%%%%%%%%%%%%%%%%%%%%%%%%%%%%%%%%%%%%%%%%%%%%%%%%%%%%%%%%%%%%%%%
\subsection{Elastic constant}
%%%%%%%%%%%%%%%%%%%%%%%%%%%%%%%%%%%%%%%%%%%%%%%%%%%%%%%%%%%%%%%%%%%%%%%%%%%%%%%%%%%%%%%%%%%%%%%%%%%

Another readily observable quantity is the elastic constant $c$
which is connected with the sound velocity.
At the phase transition, this quantity is renormalized and shows a discontinuity.
We start with the case of field-induced order at fixed temperature.
In the ordered phase ($H>H_c$),
the staggered moment is given by
\begin{align}
m^2 = -\frac{1}{2B} \left\{ a_0 \left[ 1 + \left( \frac{T}{T_0} \right)^\phi
                                         - \left( \frac{H}{H_0} \right)^2 \right] 
    + \gamma_1\epsilon+\gamma_2\epsilon^2 \right\},
\label{eqn:m}
\end{align}
which includes the magnetoelastic coupling.
When we substitute this back into the free energy (\ref{eqn:freeenergy}),
we find correction terms of order of $\epsilon^2$
which constitute a modification of the elastic constant in the ordered phase.
In the vicinity of the critical field, it is expressed as
\begin{align}
c &= c_0 -\frac{\gamma_1^2}{2B} + \frac{\gamma_2 a_0}{B} \frac{H^2-H_c^2}{H_0^2} \cr
&\simeq c_0 -\frac{\gamma_1^2}{2B} + \frac{2 \gamma_2 a_0}{B} \frac{H_c}{H_0} \frac{H-H_c}{H_0}.
\label{eqn:delC}
\end{align}
The second term gives a discontinuity in the elastic constant at $H=H_c$,
which originates from the linear magnetoelastic coupling to the strain ($\gamma_1 \epsilon m^2 $).
The $\gamma_2$ term, involving the quadratic coupling to the strain,
leads to a linear field dependence in the order phase close to the phase transition.

We again replace the unknown coefficients in eq. (\ref{eqn:delC}) by measurable quantities.
Using eqs. (\ref{eqn:gamma}) and (\ref{eqn:chi}),
we obtain the following relation in the field-dependence of the elastic constant
for a given temperature:
\begin{align}
c = c_0 -\Delta \chi c_0^2 \left[ \left( \frac{\del H_c}{\del P} \right)_T ^2 
        - \frac{1}{2}
         \left( \frac{\del^2 H_c^2}{\del P^2} \right)_T \frac{H-H_c}{H_c} \right]. 
\label{eqn:delC2}
\end{align}
We note that the discontinuity in the field dependence of the elastic constant
is negative at the transition.
This kind of discontinuous change and the linear field dependence in the ordered phase
have been measured in NH$_4$CuCl$_3$
for the elastic constant $c_{66}$ by Schmidt {\it et al}.
\cite{Schmidt,Luthi,Wolf}
From their data, they suggested that there is a certain scaling of the field dependence
of $c$ and the differential uniform susceptibility,
$\chi = \partial M / \partial H$:
\begin{align}
\chi = \frac{2 a_0 m^2}{H_0^2} + \frac{2 a_0 H}{H_0^2} \frac{\partial m^2}{\partial H} = \Delta
\chi \left(1 + 3 \frac{H-H_c}{H_c} \right).
\end{align}
However, such a scaling seems only to be attainable under the following condition:
\begin{align}
\left( \frac{\partial^2 H_c^2}{\partial P^2} \right)_T \sim
   - 6 \left( \frac{\partial H_c}{\partial P}\right)_T^2.
\end{align}
Obviously, this relation is only accidentally satisfied
and no deeper connection between $c$ and $\chi$ is found apart from the Ehrenfest relation
which relates the discontinuities of $c$ and $\chi$ at the transition.

We consider now the case when $P > P_0$ and even at zero magnetic field,
where order is possible at low enough temperature.
Then the magnetoelastic coupling results in a field dependence
of the elastic constant in the ordered phase.
Using eq. (\ref{eqn:delC}), we find
\begin{align}
c(H)-c(0) &= \frac{\gamma_2 a_0}{B} \left( \frac{H}{H_0} \right)^2 \cr
          &= \frac{c_0^2}{6} [\chi(H) - \chi(0)]
             \left( \frac{\partial^2 H_c^2}{\partial P^2} \right)_T.
\end{align}
In this case, the field dependence is more closely connected with the differential susceptibility.
The field dependence of $c$ allows us also to determine the sign of $\gamma_2$.

%%%%%%%%%%%%%%%%%%%%%%%%%%%%%%%%%%%%%%%%%%%%%%%%%%%%%%%%%%%%%%%%%%%%%%%%%%%%%%%%%%%%%%%%%%%%%%%%%%%
\subsection{Ehrenfest relations}
%%%%%%%%%%%%%%%%%%%%%%%%%%%%%%%%%%%%%%%%%%%%%%%%%%%%%%%%%%%%%%%%%%%%%%%%%%%%%%%%%%%%%%%%%%%%%%%%%%%

Finally, we turn to the Ehrenfest relations for the elastic constant.
The discontinuity of the elastic constant is given by
\begin{align}
\Delta c = -\frac{\gamma_1^2}{2B}.
\end{align}
This can be related to the discontinuities of the expansion coefficients in the following way:
\begin{align}
\frac{\Delta c}{c_0^2} = - \frac{\Delta C_p}{T_N} \left( \frac{\del T_N}{\del P} \right)_H^2
= - \Delta \alpha_1 \left( \frac{\del T_N}{\del P} \right)_H < 0,
\label{eqn:Ehrenfest-1}
\end{align}
and
\begin{align}
\frac{\Delta c}{c_0^2} = - \Delta\chi \left( \frac{\del H_c}{\del P} \right)_T ^2
= - \Delta \alpha_2 \left( \frac{\del H_c}{\del P} \right)_T < 0.
\label{eqn:Ehrenfest-2}
\end{align}
Also these relations can be in principle tested experimentally.
%In particular, the signs of the different discontinuities may be probed
%already on a qualitative level.
Since $\Delta c < 0$,
the sign of $\Delta \alpha_1$ and $\Delta \alpha_2$ is determined by
the sign of $(\frac{\partial T_N}{\partial P})_H$
and $(\frac{\partial H_c}{\partial P})_T$, respectively.

We discussed the above Ehrenfest relations for \Tl~
which shows no plateau in the magnetization curve.
In case of \N, it exhibits magnetization plateaus at 1/4 and 3/4 of the saturation moment.
\cite{Shiramura}
The recent theoretical and experimental studies revealed
that the plateaus are the consequence of successive phase transitions
driven by weakly coupled distinct dimers.
\cite{Ruegg-2004-NH4,Matsumoto-2003}
Since the each phase transition can be described by the GL theory,
the Ehrenfest relations (\ref{eqn:Ehrenfest-1}) and (\ref{eqn:Ehrenfest-2}) hold also for \N.

%%%%%%%%%%%%%%%%%%%%%%%%%%%%%%%%%%%%%%%%%%%%%%%%%%%%%%%%%%%%%%%%%%%%%%%%%%%%%%%%%%%%%%%%%%%%%%%%%%%
\subsection{Uniaxial pressure dependence of exchange interactions}
%%%%%%%%%%%%%%%%%%%%%%%%%%%%%%%%%%%%%%%%%%%%%%%%%%%%%%%%%%%%%%%%%%%%%%%%%%%%%%%%%%%%%%%%%%%%%%%%%%%

%%%%%%%%%%%%%%%%%%%%%%%%%%%%%%%%%%%%%%%%%%%%%%%%%%%%%%%%%%%%%%%%%%%%%%%%%%%%%%%%%%%%%%%%%%%%%%%%%%%
\begin{figure}[t]
\begin{center}
\includegraphics[width=7cm]{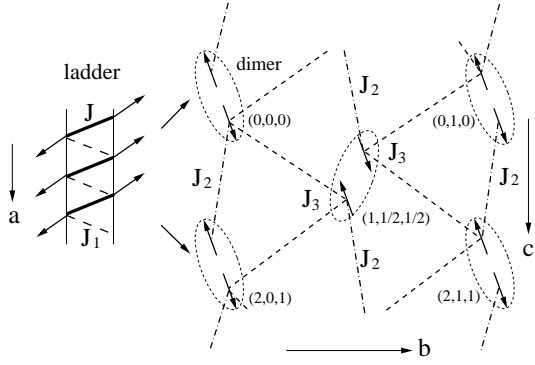}
\end{center}
\caption{
Intradimer and interdimer interactions for \Tl~and \K.
a, b, and c axes are along $(100)$, $(010)$, and $(001)$ directions, respectively.
Positions of dimers are expressed by $(x,y,z)$ in the unit of lattice spacing.
$J$ (solid line) is a intradimer interaction.
$J_1$ (dashed line) is a interdimer interaction along the ladder.
$J_2$ (dashed line) and $J_3$ (long dashed lines) are interdimer interaction
between different ladders.
}
\label{fig:Hamiltonian}
\end{figure}
%%%%%%%%%%%%%%%%%%%%%%%%%%%%%%%%%%%%%%%%%%%%%%%%%%%%%%%%%%%%%%%%%%%%%%%%%%%%%%%%%%%%%%%%%%%%%%%%%%%

%%%%%%%%%%%%%%%%%%%%%%%%%%%%%%%%%%%%%%%%%%%%%%%%%%%%%%%%%%%%%%%%%%%%%%%%%%%%%%%%%%%%%%%%%%%%%%%%%%%
\begin{figure}[t]
\begin{center}
\includegraphics[width=3.5cm]{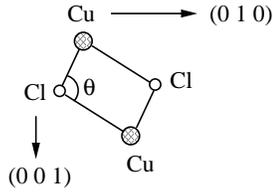}
\end{center}
\caption{
Schematic picture of exchange pathway for intradimer interaction.
}
\label{fig:dimer}
\end{figure}
%%%%%%%%%%%%%%%%%%%%%%%%%%%%%%%%%%%%%%%%%%%%%%%%%%%%%%%%%%%%%%%%%%%%%%%%%%%%%%%%%%%%%%%%%%%%%%%%%%%

In this subsection, we would like to link our discussion
to specific microscopic aspects of the isostructual compounds \Tl, \K, and \N.
For this purpose, we discuss the pressure dependence of the exchange interactions
identifying the strain in the GL free energy with some uniaxial deformations.
In particular, we discuss the sign of $(\frac{\partial T_N}{\partial P})_H$
and $(\frac{\partial H_c}{\partial P})_T$
which determine the sign of $\Delta \alpha_1$ and $\Delta \alpha_2$ at the transition,
respectively.

The compounds  \Tl~and \K~consist of two-leg ladders
which are composed of Cu$^{2+}$ and Cl$^-$ ions.
The Cu$^{2+}$ ions provide localized spin-1/2 degrees of freedom
which interact with each other through exchange paths via the Cl$^-$ ions.
Inelastic neutron scattering experiments revealed
that interladder interactions are substantial.
\cite{Kato,Cavadini-1999,Cavadini-2001,Oosawa-2002-neutron}
The system can be modeled by an isotropic Heisenberg model
with the exchange couplings given in Fig. \ref{fig:Hamiltonian}.
The two-leg ladders, which run along the $a$ axis,
are located at the corner and center of the unit cell.
The critical field is a function of intra- and interdimer interactions
\cite{Matsumoto-2002,Matsumoto-2004}
\begin{align}
g\muB H_c = \sqrt{J^2-J(J_1+J_2+2J_3)}.
\label{eqn:Hc3}
\end{align}
Since $J$ and $J_2$ are the dominant exchange interactions for \Tl,
we assume that the pressure dependence of the critical field
mainly results from the change of $J$ and $J_2$.
The critical field $H_c$ is reduced by increase of $J_2$ and decrease of $J$.
The N\'{e}el temperature $T_N$ is enhanced, accordingly.

In case of the interladder interaction $J_2$,
a shortened distance between two spins translates into an increase of the interaction.
$J_2$ is an interaction along $(2 0 1)$ direction and it would,
therefore, increase, if pressure is applied along the same direction.
In case of the intradimer interaction $J$,
the angle of the super exchange path Cu-Cl-Cu is close to 90$^\circ$ (see Fig. \ref{fig:dimer}).
The interaction strength more strongly depends on these angles
than the distance between the spins.
Application of a pressure along $(0 1 0)$ direction
increases the angle of the exchange path  away from 90$^\circ$,
and the intradimer interaction $J$ is enhanced by the pressure along $(0 1 0)$ direction.
From this, we expect
\begin{align}
&\left( \frac{\del H_c}{\del P} \right)_T > 0,~~~
\left( \frac{\del T_N}{\del P} \right)_H < 0,~~~{\rm for}~P~||~(0 1 0) \cr
&\left( \frac{\del H_c}{\del P} \right)_T < 0,~~~
\left( \frac{\del T_N}{\del P} \right)_H > 0.~~~{\rm for}~P~||~(2 0 1)
\end{align}
This determines the sign of the discontinuity of the thermal expansion $\alpha_1$ and $\alpha_2$
given in eqs. (\ref{eqn:ehrenfest-1}) and (\ref{eqn:ehrenfest-2}).
For $P~||~(0 1 0)$,  $\Delta \alpha_1 < 0$ and $\Delta \alpha_2 > 0$,
while they are opposite for $P~||~(2 0 1)$.
These features are consistent with the experimental results.
\cite{Johannsen}

%%%%%%%%%%%%%%%%%%%%%%%%%%%%%%%%%%%%%%%%%%%%%%%%%%%%%%%%%%%%%%%%%%%%%%%%%%%%%%%%%%%%%%%%%%%%%%%%%%%
\section{Summary}
%%%%%%%%%%%%%%%%%%%%%%%%%%%%%%%%%%%%%%%%%%%%%%%%%%%%%%%%%%%%%%%%%%%%%%%%%%%%%%%%%%%%%%%%%%%%%%%%%%%

Recently, it has been reported that the field-induced ordering in \Tl~
shows a weak first order phase transition by measuring ultrasonic attenuation
\cite{Sherman}
and Cl site NMR.
\cite{Vyaselev}
Measurements of other physical quantities do not so far support this finding.
While the magnetoelastic coupling, if it were strong enough,
could induce a first order transition within our theory,
we neglect this fact here and analyze the Ehrenfest relation
for second order phase transitions assuming weak coupling.

In this way, we have examined magnetoelastic effects
for the field- and pressure-induced ordered phases
based on the GL free energy introduced by eqs. (\ref{eqn:freeenergy}) and (\ref{eqn:A}).
The obtained exponents for the staggered and uniform magnetizations listed in Table I
are supported by the recent experiment
\cite{Goto}
and Monte Carlo simulations.
\cite{Nohadani}
By giving appropriate values for the GL parameters ($H_0$, $T_0$, $P_0$ and $\phi$),
we demonstrated that the GL theory reproduces the phase diagram obtained by experiments.
It enables us to have the phase boundary
as a function of temperature, pressure and magnetic field.
Moreover, our study describes the behavior of elastic constants
and expansion coefficients in the ordered phase.
Ehrenfest relations (\ref{eqn:Ehrenfest-1}) and (\ref{eqn:Ehrenfest-2}) have been found,
relating various measurable quantities.
The obtained Ehrenfest relations do not depend on details of the GL parameters,
and will help to analyze and understand experiments of magnetoelastic effects.
A rather simple discussion of the microscopic structure of spin exchange interaction
allows us to determine the sign of the discontinuities of the thermal expansion
for various uniaxial stress via the Ehrenfest relations.
A first comparison with experiments shows
that the experimentally determined coefficients give a consistent view within our GL theory.

%%%%%%%%%%%%%%%%%%%%%%%%%%%%%%%%%%%%%%%%%%%%%%%%%%%%%%%%%%%%%%%%%%%%%%%%%%%%%%%%%%%%%%%%%%%%%%%%%%%
\acknowledgments
%%%%%%%%%%%%%%%%%%%%%%%%%%%%%%%%%%%%%%%%%%%%%%%%%%%%%%%%%%%%%%%%%%%%%%%%%%%%%%%%%%%%%%%%%%%%%%%%%%%

We would like to express our sincere thanks to
N. Johannsen, S. Kimura, H. Kusunose, B. L\"{u}thi, O. Nohadani, B. Normand,
A. Oosawa, T. M. Rice, Ch. R\"{u}egg, H. Tanaka, S. Wessel, Y. Yamashita and M. Zhitomirsky 
for valuable discussions.
This work was supported by the Japan Society for the Promotion of Science (JSPS)
for Young Scientists (No. 16740197)
and the MaNEP project of the Swiss National Science Foundation.

%%%%%%%%%%%%%%%%%%%%%%%%%%%%%%%%%%%%%%%%%%%%%%%%%%%%%%%%%%%%%%%%%%%%%%%%%%%%%%%%%%%%%%%%%%%%%%%%%%%

\end{document}